\documentclass[a4paper,reqno,12pt]{amsart}
\usepackage{amssymb,euscript,bbold}
\usepackage{array}
\usepackage[T1]{fontenc}
\newcommand{\M}{\mathsf{M}}
\newcommand{\half}{\tfrac{1}{2}}
\DeclareMathOperator{\vol}{dvol}
\newcommand{\EE}{\mathbb{E}}
\newcommand{\bx}{\boldsymbol{x}}
\DeclareMathOperator{\ads}{adS}
\newcommand{\D}{\mathsf{D}}
\newcommand{\RR}{\mathbb{R}}

\renewcommand{\Sp}{\mathrm{Sp}}
\newcommand{\Spin}{\mathrm{Spin}}
\newcommand{\U}{\mathrm{U}}
\newcommand{\SU}{\mathrm{SU}}
\newcommand{\SO}{\mathrm{SO}}
\renewcommand{\d}{\partial}
\newcommand{\su}{\mathfrak{su}}
\newcommand{\CP}{\mathbb{CP}}
\newcommand{\TT}{\mathbb{T}}
\begin{document}

\title[Near-horizon geometries]{Near-horizon geometries of
supersymmetric branes}
\author{Jos\'e M Figueroa-O'Farrill}
\address[]{\begin{flushright}Department of Physics\\
Queen Mary and Westfield College\\
Mile End Road\\
London E1 4NS, UK\end{flushright}}
\email{j.m.figueroa@qmw.ac.uk}
\date{July 13, 1998}
\begin{abstract}
This is the written version of my talk at {\em SUSY '98\/}.  It
presents a geometric characterisation of the allowed near-horizon
geometries of supersymmetric branes.  We focus primarily on the
$\M2$-brane, but results for other branes (e.g., the $\D3$-brane) are
also presented.  Some new examples are discussed.
\end{abstract}
\maketitle


\section{Motivation}

There has been a lot of recent activity on testing and generalising
the conjectured duality \cite{Malda,GKP,WittenAdS} between
(the 't~Hooft limit of) superconformal field theories and (the
supergravity limit) of superstring and $\M$-theories.

In Maldacena's argument one decouples the gauge theory on the brane
from the bulk and studies the theory near the brane horizon.  For the
maximally supersymmetric theories, the near-horizon geometry is always
a product of an anti~de~Sitter space and a round sphere; but one can
obtain theories with less supersymmetry on branes with non-spherical
near-horizon geometries; that is, by replacing the sphere by some
other manifold $X$.  It is therefore interesting to determine the
possible $X$ which can appear and to explore their dual field
theories.  In this talk I will restrict myself to presenting a
geometric characterisation of those $X$ which give rise to
supersymmetric theories, but make no attempt to relate the
near-horizon geometries of the branes to the field theories on them.
This aspect of the work will appear in \cite{AFHS}.

Although this work has certainly been motivated by the Maldacena
conjecture, it is not based on it and as a result I have made no
attempt in this talk to compare these results with the increasing
number of papers which discuss non-spherical near-horizon geometries
in this context.  This will be remedied in our forthcoming paper
\cite{AFHS}.

The talk is organised as follows.  I shall first discuss the
near-horizon geometries of the elementary branes, focusing on the
$\M2$-brane.  We will then see how to generalise the solutions to
obtain a different near-horizon geometry with less supersymmetry.  The
possible supersymmetric geometries will be characterised and examples
will be given.  I will conclude with some remarks and directions for
future work.

\section{Near-horizon geometries of supersymmetric $p$-branes}

We start by briefly reviewing the near-horizon geometry of the
membrane solution to $D{=}11$ supergravity.  We then briefly
generalise this to other supersymmetric $p$-branes.

\subsection{Eleven-dimensional supergravity}

Eleven-dimensional supergravity consists of the following fields
\cite{Nahm,CJS}: a Lorentzian metric $g$, a closed $4$-form $F$ and a
gravitino $\Psi$.  By a supersymmetric vacuum I will mean any solution
of the equations of motion with $\Psi=0$ for which the supersymmetry
variation $\delta_\varepsilon \Psi=0$, as an equation on the spinor
parameter $\varepsilon$, has solutions.  Being linear in
$\varepsilon$, the solution space will have dimension $n$.  The
eleven-dimensional spinorial representation is $32$-dimensional and
real, so there at most $32$ linearly independent solutions.  An
important physical invariant of a supersymmetric vacuum is the
fraction $\nu \equiv n/32$ of the supersymmetry that the solution
preserves.  For example, $F=0$ and $g$ the flat metric on
eleven-dimensional Minkowski spacetime is a supersymmetric vacuum with
$\nu=1$---that is, it is maximally supersymmetric.  Other maximally
supersymmetric vacua are $\ads_4\times S^7$ and $\ads_7 \times S^4$
with $\star F$ or $ F$ having quantised flux on the $S^7$ or $S^4$,
respectively.

Eleven-dimensional supergravity has four types of elementary
solutions with $\nu=\half$: the $pp$-wave and the Kaluza--Klein
monopole, about which I will have nothing else to say in this talk,
and the elementary brane solutions: the {\em electric\/} membrane
\cite{DS2brane} and the {\em magnetic\/} fivebrane \cite{Guven}.  I
will focus primarily on the membrane, since as we will see, it will
allow for a much richer near-horizon geometry.

\subsection{The supermembrane}

The elementary membrane solution has the following form
\cite{DS2brane}:
\begin{align}\label{eq:M2}
g &= H^{-\frac23}\, g_{2+1} + H^{\frac13} g_{8}\\
F &= \pm \vol_{2+1} \wedge dH^{-1}\notag~,
\end{align}
where $g_{2+1}$ and $\vol_{2+1}$ are the metric and volume form on the
three-dimensional Minkowskian worldvolume of the membrane $\EE^{2,1}$;
$g_8$ is the metric on eight-dimensional euclidean transverse space
$\EE^8$; and $H$ is a harmonic function on $\EE^8$.  For example, if
we demand that $H$ depend only on the transverse radial coordinate $r$
on $\EE^8$, we then find that
\begin{equation}\label{eq:H(r)}
H(r) = 1 + \frac{\alpha}{r^6}
\end{equation}
is the only solution with $\lim_{r\to\infty} H(r) = 1$.  This
corresponds to one of more coincident membranes at $r=0$.  Flux
quantisation of $F$ implies that $\alpha$ is quantised in units of
$\ell_{11}^6$, with $\ell_{11}$ the eleven-dimensional Planck length.

I should remark that more general $H$ are possible, corresponding to
parallel membranes localised at the singularities of $H$.  In fact, we
can take $H(\bx)$ to be an arbitrary harmonic function on $\EE^8$ with
suitable asymptotic behaviour $H(\bx) \to 1$, say, as
$|\bx|\to\infty$.

Although the membrane solution \eqref{eq:M2} with $H$ given by
\eqref{eq:H(r)} only preserves $\nu = \half$ of the supersymmetry, it
nevertheless interpolates between two maximally supersymmetric
solutions: flat $\EE^{10,1}$ infinitely far away from the membrane,
and $\ads_4 \times S^7$ near the brane horizon
\cite{GibbonsTownsend,DuffGibbonsTownsend}.  Let us see this in more
detail.

\subsection{Near-horizon geometry}

Notice that the metric on the transverse space is given by
\begin{equation}\label{eq:EScone}
g_8 = dr^2 + r^2\,g_S~,
\end{equation}
where $g_S$ is the metric on the unit sphere $S^7 \subset \EE^8$.  In
the near-horizon limit,
\begin{equation*}
\lim_{r\to 0} H(r) \sim \frac{\alpha}{r^6}~,
\end{equation*}
so that the membrane metric becomes
\begin{equation*}
g = \alpha^{-\frac23} r^4 g_{2+1} + \alpha^{\frac13} r^{-2} dr^2 +
\alpha^{\frac13} g_S~.
\end{equation*}
The last term is the metric on a round $S^7$ of radius
$R=\alpha^{\frac16}$; whereas the first two combine to produce the
metric on $4$-dimensional anti~de~Sitter spacetime with ``radius''
$R_{\ads} = \half R$ in Bertotti-Robinson coordinates:
\begin{equation*}
g_{\ads} = R^2_{\ads} \left[ \frac{du^2}{u^2} + \left(
\frac{u}{R_{\ads}} \right)^2 \frac{g_{2+1}}{R_{\ads}^2} \right]~,
\end{equation*}
with $u = \frac14 r^2/R_{\ads}$.

\subsection{Other branes}

Similar considerations apply to other branes.  The magnetic fivebrane
solution interpolates between Minkowski spacetime asymptotically far
away from the brane to $\ads_7 \times S^4$ near the horizon.  We will
see, however, that the allowed near-horizon geometries for the
fivebrane is much more rigid than for the membrane above.

We can also consider $\D3$-branes in type IIB
\cite{HorowitzStrominger}, whose metric is given by
\begin{equation}\label{eq:D3}
g = H^{-\half}\, g_{3+1} + H^{\half}\, g_6~,
\end{equation}
where $g_{3+1}$ is the metric on the Minkowski worldvolume of the
brane $\EE^{3,1}$ and $g_6$ is the euclidean metric on the transverse
$\EE^6$.  The self-dual $5$-form $F$ has (quantised) flux on the unit
transverse five-sphere $S^5\subset \EE^6$.  $H$ is again harmonic and,
if radially symmetric, is given by
\begin{equation}\label{eq:H(r)D3}
H(r) = 1 + \frac{\beta}{r^4}~,
\end{equation}
where now $\beta$ is quantised in units of $\ell_{10}^4$, with
$\ell_{10}$ the ten-dimensional Planck length.  The near-horizon
geometry (see, for example, \cite{Malda}) is given now by
\begin{equation*}
\ads_5 \times S^5~,
\end{equation*}
where both the radii of anti~de~Sitter space and the sphere are equal
to $\beta^{\frac14}$.

Generally there are supersymmetric $p$-branes in $D$ dimensions with
near-horizon geometry (we assume $D-p>4$)
\begin{equation*}
\ads_{p+2} \times S^{D-p-2}~.
\end{equation*}
These solutions are all maximally supersymmetric.  Sacrificing some
(but not all!) of the supersymmetry, one can obtain $p$-branes
with more interesting near horizon geometries.  This possibility had
been envisaged, for example, in \cite{DLPS,CCDAFFT}.

\section{Non-spherical horizons of generalised $p$-branes}

In this section I consider the near-horizon geometries of generalised
$p$-branes preserving a smaller fraction of the supersymmetry, and I
will answer the question posed at the start of this talk about which
are the allowed near-horizon geometries.

\subsection{Generalised $p$-branes}

Let us look for solutions whose near-horizon geometries are of the
form
\begin{equation*}
\ads_{p+2} \times M^{D-p-2}~,
\end{equation*}
where $M$ is a compact Einstein manifold with cosmological constant
$\Lambda = D-p-3$, just as for the standard unit sphere in
$\EE^{D-p-1}$.  This is necessary for solving the supergravity
equations of motion.

Such a near-horizon geometry arises from generalised $p$-branes whose
transverse space is the (deleted) metric cone $C(M)$ of $M$.
Topologically, $C(M) \cong \RR^+ \times M$, where $\RR^+$ is the
half-line, with parameter $0 < r < \infty$.  Geometrically, the cone
metric is not a product, but rather
\begin{equation*}
g_C = dr^2 + r^2 \, g_M~,
\end{equation*}
where $g_M$ is the Einstein metric on $M$.

If $M= S^{D-p-2}$ is the sphere, then $C(M) = \EE^{D-p-1} \setminus
\{0\}$ is Euclidean space with the origin deleted.  In this case (and
only in this case), the metric extends smoothly to the apex of the
cone: the point $r=0$.  More generally, if $(M,g_M)$ is Einstein with
$\Lambda = \dim M - 1$, then $C(M)$ is Ricci-flat with a conical
singularity; although this does not mean that the brane solution is
necessarily singular.

For example, if we substitute the euclidean transverse metric $g_8$ by
such a $g_C$ in the $\M2$-brane solution \eqref{eq:M2}, we obtain a
solution of the supergravity equations of motion with the same $H$ as
in \eqref{eq:H(r)} \cite{CCDAFFT}. Similarly, substituting $g_6$ for
an appropriate $g_C$ in the $\D3$ solution \eqref{eq:D3} yields a
solution of the type IIB supergravity equations of motion with the
same $H$ as in \eqref{eq:H(r)D3} \cite{Kehagias}.

Not every solution will be supersymmetric, however.  Demanding that
the solution preserve some supersymmetry implies that $M$ should admit
Killing spinors, satisfying:
\begin{equation*}
\nabla^{(M)}_m \psi = \pm\half \Gamma_m \,\psi~,
\end{equation*}
where the sign depends on the sign of $F$ in the solution.  It is easy
to show that this equation is equivalent to
\begin{equation*}
\nabla^{(C)}_M \psi = 0~,
\end{equation*}
where $\nabla^{(C)}$ is the spin connection on the metric cone
$C(M)$.  In other words, $M$ admits real Killing spinors if and only
if its cone $C(M)$ admits parallel spinors \cite{Baer}.

In what follows we will make the simplifying assumption that $M$ (and
hence $C(M)$) is simply-connected.  Quotients, be they smooth or
orbifolds, have to be analysed case by case, whereas for the purposes
of this talk I am interested only in the generic case.

Simply-connected riemannian manifolds admitting parallel spinors are
classified by their holonomy group \cite{Wang}.  Because $C(M)$ is
Ricci-flat, we know that it cannot be a locally symmetric space.
Moreover, by a theorem of Gallot \cite{Gallot}, the holonomy group
acts on $C(M)$ irreducibly unless $C(M)$ is flat, in which case $M$ is
the round sphere. Therefore, we need only consider irreducible
holonomy groups of manifolds which are not locally symmetric.  In
other words, those in Berger's table (see, e.g.,
\cite{Besse,Salamon}).

Of those, the ones which admit parallel spinors are given in the
following table, which also lists the number $N$ (or
$(N_L,N_R)$ in even $D$) of linearly independent parallel spinors.

\begin{table}[h!]
\centering
\setlength{\extrarowheight}{3pt}
\begin{tabular}{|>{$}c<{$}|>{$}c<{$}|c|>{$}c<{$}|}\hline
D& \text{Holonomy} & Geometry& N\\[3pt]
\hline
\hline
4k+2 & \SU(2k+1) & Calabi--Yau & (1,1)\\
4k & \SU(2k) & Calabi--Yau & (2,0)\\
4k & \Sp(k) & hyperk\"ahler & (k+1,0)\\
7 & G_2 & exceptional & 1\\
8 & \Spin(7)  & exceptional & (1,0)\\[3pt]
\hline
\end{tabular}
\vspace{8pt}
\caption{$D$-dimensional manifolds admitting parallel spinors}
\label{tab:berger}
\end{table}

If $M$, and hence $C(M)$, were not simply-connected, then the
restricted holonomy group would be contained in the above table.
However a spinor which is invariant under the restricted holonomy
group need not be parallel, because it may still transform
nontrivially under parallel transport along noncontractible loops.
Therefore the number of parallel spinors in $C(M)$ will be at most the 
number shown in the table.

\subsection{Near-horizon geometries}

What can one say about the near-horizon manifold $M$ when its cone
$C(M)$ is one of the manifolds in Table \ref{tab:berger}?  We will see
that $M$ admits some interesting geometric structures built out of the
natural objects present in $C(M)$.

First and foremost, $C(M)$ is a cone.  This means that it has a
privileged vector field, the Euler vector $\xi \equiv r\d_r$, which
generates the rescaling diffeomorphisms.  In addition, the cone has
reduced holonomy, whence the holonomy principle guarantees the
existence of certain parallel tensors.  Out of these tensors and the
Euler vector, one can construct interesting geometric structures on
$M$, identified with $\{1\}\times M \subset C(M)$.

For example, if $C(M)$ has $\Spin(7)$ holonomy, then it has a
parallel self-dual $4$-form $\Omega$---the Cayley form.  We can define
a $3$-form $\phi$ on $M$ by contracting the Euler vector into the
Cayley form:
\begin{equation*}
\phi \equiv \imath(\xi)\cdot\Omega\quad\text{so that}\quad
\Omega = dr\wedge \phi + \star \phi~.
\end{equation*}
Because $\Omega$ is parallel in $C(M)$, it follows that
$\nabla\phi = \star\phi$ on $M$.  In other words \cite{Gray-WH},
$M$ has {\em weak $G_2$ holonomy\/}.  The converse also holds, so that
$M$ has weak $G_2$ holonomy if and only if its metric cone has
$\Spin(7)$ holonomy \cite{Baer}.

Similarly, if $C(M)$ has $G_2$ holonomy, then it possesses a parallel
$3$-form $\Phi$, the associative $3$-form, which together with its
Hodge dual $4$-form $\Tilde\Phi \equiv \star\Phi$ generate the ring of
parallel forms.  Contracting the Euler vector into the associative
$3$-form $\Phi$ obtains a $2$-form
\begin{equation*}
\omega \equiv \imath(\xi)\cdot\Phi~.
\end{equation*}
This form defines an almost complex structure $J$ on $M$ by
\begin{equation*}
\langle X,J\,Y\rangle = \omega(X,Y)~.
\end{equation*}
Because $\Phi$ is parallel in $C(M)$, $J$ obeys $\nabla_X J (X) = 0$
for any vector field $X$, but $\nabla_X J\neq 0$.  In other words
\cite{GrayNK}, $M$ is {\em nearly K\"ahler\/} but not K\"ahler.  Again
there is a converse to this, so that a manifold is nearly K\"ahler but
non-K\"ahler if and only if its metric cone has $G_2$ holonomy
\cite{Baer}.

Similarly it is possible to recognise the geometric structures for the
hyperk\"ahler and Calabi--Yau cases.

Every parallel complex structure $I$ on $C(M)$ gives rise to a
{\em Sasaki\/} \cite{Sasaki} structure on $M$:
\begin{itemize}
\item a unit norm Killing vector $X$, which in this case is given by
      $I\,\xi$;
\item a dual $1$-form $\theta = \langle X,-\rangle$;
\item a $(1,1)$-tensor $T = - \nabla X$ satisfying, for all vector
      fields $V,W$,
\begin{equation*}
\nabla_V T(W) = \langle V,W \rangle X - \theta(W)\, V~.
\end{equation*}
\end{itemize}

In fact, it is easy to see that $C(M)$ is K\"ahler if and only $M$ is
Sasaki.  Because $C(M)$ is Calabi--Yau if and only if it is Ricci-flat
and K\"ahler, it follows that $C(M)$ is Calabi--Yau if and only if $M$
{\em Sasaki--Einstein\/}; that is, Sasaki and Einstein
\cite{FK5,FK7-2,Baer}.  Similarly it is shown in \cite{FK7-2,Baer,BGM}
that $C(M)$ is hyperk\"ahler if and only if $M$ is {\em 3-Sasaki\/},
so that it has three Sasaki structures $(X_i,\theta_i,T_i)$ for
$i=1,2,3$ obeying a series of identities derived from the fact that
the Killing vectors $X_i$ satisfy an $\su_2$ Lie algebra.

A good review of the geometry of manifolds admitting Killing spinors
is given in \cite{BFGK}.  For the latest word on Sasaki structures,
see \cite{BoGa}.

\subsection{New supersymmetric horizons}

Let us now apply the preceding geometrical analysis to the case at
hand.  If the transverse space to a supersymmetric $p$-brane in $D$
dimensions is a simply-connected metric cone $C(M)$, then the
near-horizon geometry is of the form $\ads_{p+2} \times M^d$, where
$d=D-p-2$.  Every such solution will preserve a fraction $\mu$ of the
supersymmetry {\em relative to the round sphere\/}, which is maximally
supersymmetric.  Changing the orientation of $M$ yields a fraction
$\Bar\mu$.  The possible geometries together with the fractions are
listed in the following table.

\begin{table}[h!]
\centering
\setlength{\extrarowheight}{3pt}
\begin{tabular}{|>{$}c<{$}|c|>{$}c<{$}|}\hline
d & Geometry of M & (\mu, \Bar\mu)\\
\hline
\hline
\text{any} & round sphere & (1,1)\\
7 & weak $G_2$ holonomy & (\frac18,0)\\
  & Sasaki--Einstein & (\frac14,0)\\
  & 3-Sasaki & (\frac38, 0)\\
6 & nearly K\"ahler & (\frac18,\frac18)\\
5 & Sasaki--Einstein & (\frac14,\tfrac14)\\[3pt]
\hline
\end{tabular}
\vspace{8pt}
\caption{Possible near-horizon geometries of supersymmetric branes}
\label{tab:geometries}
\end{table}

Notice in particular that for $d=4$, corresponding to the $\M5$-brane,
the only simply-connected geometry is spherical.

\section{Examples}

We now take a look at examples of some of these manifolds.  The
homogeneous examples were studied intensively in the eighties in the
context of Kaluza--Klein compactifications of supergravity theories
\cite{CastellaniRomansWarner,DNP}.  However many non-homogeneous
examples have recently been constructed, especially in $7$ dimensions
which, as can be gleaned from Table \ref{tab:geometries}, is the
richest dimension.  This corresponds to the near-horizon geometries of
generalised $\M2$-branes \cite{DLPS}.

\subsection{Seven dimensions}

Consider the generalised mambrane solution \eqref{eq:M2} and
\eqref{eq:H(r)} and replace the transverse space $\EE^8$ by a cone
over a seven-dimensional Einstein manifold $M$.  We can take $M$ from
the following (incomplete) list:
\begin{itemize}
\item 3-Sasaki ($\nu = \tfrac3{16}$)
\begin{itemize}
\item $\SU(3)/\mathrm{S}\left(\U(1) \times
\U(1)\right)$
\item $N_{010}$ \cite{CastellaniRomans}
\item A new infinite toric family recently studied in
\cite{Bielawski} and \cite{BGMR}
\end{itemize}
\item Sasaki--Einstein ($\nu = \tfrac18$)
\begin{itemize}
\item $M_{pqr}$ \cite{WittenMpqr,CastellaniRomansWarner}
\item Circle bundles over K\"ahler 3-folds (see, e.g., \cite{BoGa})
\begin{itemize}
\item[$\circ$] $\CP^1\times \CP^1\times \CP^1$
\item[$\circ$] $\CP^1 \times \CP^2$
\item[$\circ$] $\CP^3$
\item[$\circ$] $\SU(3)/\TT^2$
\item[$\circ$] $\mathrm{Gr}(2|5)$
\end{itemize}
\end{itemize}
\item weak $G_2$ holonomy ($\nu = \tfrac1{16}$)
\begin{itemize}
\item Any squashed \cite{GalickiSalamon} 3-Sasaki
manifold (e.g., the squashed $7$-sphere)
\item $\SO(5)/\SO(3)$ \cite{CastellaniRomansWarner}
\item $N_{pqr}$ \cite{CastellaniRomans}
\item The Aloff--Wallach spaces \cite{AloffWallach}
\end{itemize}
\end{itemize}

Let us remark that the manifolds listed here comprise infinitely many
different homotopy types.  In fact, this already happens just for the
Aloff--Wallach spaces, which in addition contain exotic differentiable
structures \cite{KreckStolz}.  The infinite toric family of
\cite{Bielawski,BGMR} consists of non-homogeneous examples which
contain all the possible rational homotopy types allowed for 3-Sasaki
7-dimensional manifolds: $b_2 = b_5$ are the only nonzero Betti
numbers, and there are examples with arbitrary $b_2$ \cite{BGMR}.  The
cones over this toric family form a subclass of a larger class of
toric hyperk\"ahler eight-dimensional manifolds \cite{BD}, which was
also studied in \cite{GGPT} in the context of intersecting branes.  In
fact this work was reported on by Gibbons in this same conference a
year ago \cite{GibbonsSUSY97}.

\subsection{Six dimensions}

The maximally supersymmetric solution consists of taking $S^6$, but
solutions with $\tfrac18$ as much supersymmetry are possible by taking
$M$ to be a nearly K\"ahler manifold.  Examples of nearly K\"ahler
six-dimensional manifolds are
\begin{itemize}
\item $\CP^3 \cong \SO(5)/\U(2)$ with the natural $\SO(5)$-invariant
      metric;
\item The complex flag manifold $F(1,2|3) \cong
      \U(3)/[\U(1)\times\U(1)\times\U(1)]$ with the natural
      $\U(3)$-invariant metric;
\item $S^3 \times S^3$ with the metric making into a riemannian
      3-symmetric space (see, e.g., \cite{Grunewald});
\end{itemize}

Six-dimensional nearly K\"ahler manifolds have many properties similar
to Calabi--Yau 3-folds.  For example, non-K\"ahler such manifolds have
vanishing first Chern class \cite{GrayNK2}.

\subsection{Five dimensions}

These manifolds can be understood as the near-horizon geometries of
$\D3$-branes in type IIB superstring theory in ten dimensions.  The
maximally supersymmetric geometry is $S^5$, but geometries with
$\tfrac14$ as much supersymmetry are possible by taking $M$ to be a
Einstein--Sasaki manifold.  Those for which the action of the Killing
vector $X$ in the Sasaki structure fibres, are circle bundles over the
del Pezzo surfaces $P_k$ with $3\leq k\leq 8$ with K\"ahler--Einstein
metric \cite{TianYau} or $\CP^1\times\CP^1$ \cite{FK5,BoGa,Kehagias}.

\section{Conclusions}

Geometry and supersymmetry have always benefitted from a healthy and
fruitful relation and the latest developments in superstring theory
are no exception.  I hope to have exhibited convincing evidence of the
richness of near-horizon geometries for supersymmetric branes.
Conjecturally, these geometries should each have a dual quantum field
theory and the obvious next step is to investigate this duality in
detail to see what can be learned about the quantum field theory from
the near-horizon geometry (and eventually even viceversa).  In our
forthcoming work \cite{AFHS} we will report on a first step in this
direction.

We have made a simplifying assumption about the topology of the
near-horizon geometry; namely that it is simply-connected.  Clearly it
would be very interesting to consider finite quotients of these spaces
and obtain other geometries with less supersymmetry.

Another interesting problem is the following.  The toric examples of
$7$-dimensional $3$-Sasaki manifolds considered above were constructed
via a quotient construction \cite{BGM3SQ} akin to the hyperk\"ahler
quotient \cite{HKLR}.  Does the $3$-Sasaki quotient have a direct
supersymmetric origin? 

Finally it would be interesting to investigate the effect of
superstring dualities on the near-horizon geometries; and in
particular to see if these generalised branes with less supersymmetry
are dual to configurations of intersecting (static or not) elementary
branes.

I hope to report on these and other related problems at a future
date.

\section*{Acknowledgement}

It is my pleasure to thank Herbi Dreiner and the other organisers of
{\em SUSY '98\/} for a pleasant and stimulating conference.  I would
like to thank the following people for useful conversations: Bobby
Acharya, Krzysztof Galicki, Jerome Gauntlett, Chris Hull, Bill Spence
and Sonia Stanciu.  I would also like to thank Arkady Vaintrob for
insisting, although he might not remember this, that Sasaki structures
would come in handy some day.  Last but certainly not least, I would
like to ackowledge the financial support of the EPSRC, under grant
GR/K57824.

%
%

\end{document}